\documentstyle[psfig,preprint,tighten,floats,aps]{revtex}

\def\ut#1{\rlap{\lower1ex\hbox{$\sim$}}{#1}}
\def\l{\ell}

\def\bF{{\bf F}}

\def\bA{{\bf A}}

\def\pb#1{\rlap{\lower1.5ex\hbox{$\longleftarrow$}}{#1}}
\def\dpb#1{\rlap{\lower1.5ex\hbox{$\Longleftarrow$}}{#1}}
\def\spb#1{\rlap{\lower1.0ex\hbox{$\leftarrow$}}{#1}}
\def\sdpb#1{\rlap{\lower1.0ex\hbox{$\Leftarrow$}}{#1}}
\def\d{{\rm d}}

\def\ba{\begin{eqnarray}}
\def\ea{\end{eqnarray}}
\def\be{\begin{equation}}
\def\ee{\end{equation}}
\def\={\mathrel{\widehat\mathalpha{=}}}
\def\puto#1{\rlap{\raise.5ex\hbox{\char'27}}{#1}}

\preprint{\vbox{\baselineskip=12pt
\rightline{GTC-99/12-1}
\rightline{gr-qc/9912032}
}}

\begin{document}
\draft
\title{
Mass of Colored Black Holes}
\author { 
Alejandro\ Corichi\thanks{E-mail: 
corichi@nuclecu.unam.mx} and Daniel Sudarsky\thanks{E-mail:
sudarsky@nuclecu.unam.mx}
}

\address{Instituto de Ciencias Nucleares\\
Universidad Nacional Aut\'onoma de M\'exico\\
A. Postal 70-543, M\'exico D.F. 04510, M\'exico.}

\maketitle

\begin{abstract}
New results pertaining to colored static black hole 
solutions to the
Einstein-Yang-Mills equations  are obtained.
The isolated horizons framework is used to define
the concept of Hamiltonian Horizon Mass of the
black hole. 
An unexpected relation
between the ADM  and Horizon masses of the black hole solution
and the 
ADM mass of the corresponding Bartnik-McKinnon soliton 
is found. 
These results can be generalized to other non-linear theories
and they suggest a general testing bed for the instability of
the corresponding hairy black holes. 
\end{abstract}

The laws of BH mechanics
\cite{2} and an essentially complete set of  uniqueness 
theorems\cite{chru,mh} have led
to a rather complete understanding of the properties of
stationary solutions in  Einstein-Maxwell-Scalar Field  
theories \cite{mh}. 
However, with
the appearance of new, hairy, black holes
in other theories with non-linear matter couplings, 
such as non-abelian gauge theories,
several new issues arose. 
In particular, the
physical significance of the discrete families of colored 
black holes in, say, the Einstein-Yang-Mills (EYM) system
still remains somewhat obscure.  

One issue that has been considered in that context is the relation
that might exist between the existence of regular 
static, solitonic solutions and  `hairy' black hole solutions.  
(This issue has been considered for example in \cite{bizon2}
from heuristic and dimensional arguments.)

The purpose of this letter is to shed new light on these issues and
to establish unexpected relations between the two classes of solutions 
that coexist in one theory. We will specifically address the case of
static spherically symmetric solutions (SSS) to the 
EYM equations, but it will be apparent that the methodology can be extended 
straightforwardly to other cases. 
The basic input that we use to arrive at these new results comes
form the recently formulated ``Isolated Horizon" (IH) 
framework \cite{ack,abf2}.
In particular, two main issues are studied. First,
by making a crucial use of
the Hamiltonian formulation for Isolated black holes we 
define the Hamiltonian Horizon Mass (HHM)
of SSS black holes in EYM theory.
We then use this expression to make
the main observation of this
letter, namely, to show that this quasi-local definition 
together with
some basic properties of Hamiltonian Mechanics lead us to a formula 
relating HHM and ADM mass of the colored
BH solutions with the ADM mass of the Solitons of 
the theory. We conclude that the 
positivity of the `total energy' spectrum of the colored black 
holes is related to their instability.

These results are quite surprising, because the IH formalism
was developed to extend the notion of black holes
to situations where radiation is present --and goes
out to infinity-- and one might
have not expected to obtain new results already  
in the static sector of the theory.

In this note we focus our attention to the
Einstein-Yang-Mills system defined by the action,
\be S_{\rm EYM} (\bA) =-\frac{1}{16\pi}\int_{{\cal M}}\sqrt{-g} 
[R+ \bF^i_{ab}\bF_i^{ab}]\d^4x\, ,\label{dil:act} 
\ee
where the abstract indices $a,b,\ldots$ denote space-time objects and the 
indices $i,j,\ldots$ are internal indices in the Lie algebra of the gauge 
group $G$. In this letter we restrict out attention to $G=SU(2)$.
The field strength $\bF$ is given by 
$\bF^i_{ab}=2\nabla_{[a}\bA^i_{b]}+
{\epsilon^i}_{jk}\bA_a^j\bA_b^k$, that is, the 
curvature of the Lie algebra valued one form $\bA_a^i$.
We note that this theory has a characteristic scale given by a combination
 of Newton's constant
 and the  coupling constant of the gauge theory. In the present work that
 scale has been absorbed in the dimension of the coordinates.
The equations of motion that follow from  $S_{\rm EYM}$, together with the 
Bianchi identity for the YM sector, are:
\ba 
D_a {}\bF^{iab} = 0,\quad\quad
D_{[c}\bF_{ab]}=0,\label{dil:eom1}\\
R_{ab}= 2\left(\bF^i_{ac}{\bF_{ib}}^c -\frac{1}{4}
g_{ab}\bF^2  \right),
\label{dil:eom3}
\ea
where $\bF^2=\bF^i_{ab}\bF_i^{ab}$, and $D_a$ is the generalized 
covariant derivative defined by $\bA$. The dual field tensor is
given by ${}^*\bF_{ab}=\frac{1}{2}{\epsilon_{ab}}^{cd}\bF_{cd}$, 
where ${\epsilon_{abcd}}$ is the volume form associated with the metric.
We can define gauge invariant quantities, for any two-sphere
$S$ as follows,
\be
Q_S:=\frac{1}{4\pi}\oint_{S}|{}^*\bF|, \qquad
P_S:=\frac{1}{4\pi}\oint_{S}|\bF| \label{QoP},
\ee
where by the two form $|\bF|$ we mean the following: 
take $\epsilon^{ab} $, the area two form associated with the 
2-sphere $S$ and define $f^i = \bF^i_{ab} \epsilon^{ab}$. 
Then $|\bF|_{ab} =\sqrt{\sum (f^i)^2} \epsilon_{ab}$.

For the results of this letter, 
the details of the IH analysis in EYM are not
necessary. Thus, we only refer to those results
of the Hamiltonian formulation that we use 
for our discussion (details can be found in \cite{ac:ds}).
In the Einstein-Yang-Mills case the total Hamiltonian
consists of a bulk term and \textit{two} surface terms,
one at infinity and the other at
the isolated horizon.  As usual, the bulk term is a linear combination
of constraints and the surface term at infinity yields the ADM energy.
In a rest-frame adapted to the horizon it is then natural to identify
the surface term at the horizon $\Delta$ as the Horizon Mass, 
$M_\Delta$.  

We Consider a foliation of the given
space-time region ${\cal M}$ by a 1-parameter family of (partial)
Cauchy surfaces $M_t$, each of which extends from the isolated horizon
$\Delta$ to spatial infinity $i^o$ . Then,
the Hamiltonian $H_t$ generating evolution along the properly
normalized vector field $t^a$ that approaches $l^a$,
 the null generator of the horizon $\Delta$, is given by:
\ba \label{ham}
H_t &=& \int_M {\rm constraints}\,\,- M_{\rm ADM}  \nonumber\\
&+& \oint_{S_\Delta} \left(\frac{\mu^{-1}}{4\pi} \Psi_2 \right) 
{}^2\epsilon + |(\bA\cdot l)|{Q}_\Delta + V\, ,
\ea
with $\mu$  a normalization factor for the vector field $l^a$
generating the horizon chosen as in \cite{abf2},
$|\bA\cdot l |$  the norm of $\bA^i_al^a$ and $Q_\Delta$  the
charge as defined by (\ref{QoP}) evaluated at the horizon $\Delta$.
As already mentioned, we identify
the surface term at $S_\Delta$ as the HHM $M_\Delta$
of the isolated horizon. 
The quantity $V$ is constant over the region of
phase space being considered in (\ref{ham}), and arises only in
the Hamiltonian framework.
Using some identities that follow from the IH boundary conditions
\cite{abf2}, one can define
the surface gravity in terms of the
Newman-Penrose component $\Psi_2$ and $\mu$. If we now fix
the value of $\Phi:=|\bA\cdot \l|$ on
$\Delta$ to coincide with the value it takes in the
(Abelian) family of static solutions,
we can cast $M_\Delta$ in a more familiar form:
\be 
M_\Delta = \frac{1}{4\pi}\, \kappa a_\Delta\, +\, \Phi Q_\Delta 
+V,\label{smarr}
\ee
where $a_\Delta$ is the horizon area.
Thus, we obtain a Smarr formula for the mass of the black hole. The
quantity $V$ now becomes dependent on the intrinsic --absolutely
conserved-- parameters of the black hole horizon
such as $Q_\Delta$, $P_\Delta$ and $a_\Delta$, since we
now consider the full Isolated Horizon phase space (with
all posible values of  $Q_\Delta$, $P_\Delta$, $a_\Delta$).
Its functional
dependence is not arbitrary but is restricted by the mass
variation formula coming from the IH formalism,
\be
\delta M_\Delta=\frac{1}{8\pi}\,\kappa\, \delta a_\Delta
+ \Phi\,\delta Q_\Delta\, . \label{11}
\ee  
This equation is a necessary condition for the existence of
a consistent Hamiltonian framework, and is sufficient to
determine $V$.

Even when Equation (\ref{smarr}) resembles the
Smarr formula for static space-times, the meaning of 
various symbols in the equation is somewhat different.
Since an isolated horizon need not be a Killing horizon, in general
$M_\Delta$ does \text{not} equal the ADM mass, nor do $\kappa$ or
$\Phi$ refer to a Killing field. Given that
the constraints are satisfied in any solution, the bulk term in
(\ref{ham}) vanishes as well.  Hence, in this case, 
$H_t=M_\Delta - E^{\rm
ADM}$,  the `total energy'
in the space-time, i.e., the energy available 
to be radiated, as is clear from the fact that,
in a dynamical process, 
$\delta H_t=\delta E^{\rm Rad}_{\infty}$ \cite{abf2}.
Finally, as emphasized in \cite{abf2}, the matter
contribution to the mass formula (\ref{smarr}) is subtle: while it does
not include the energy in radiation outside the horizon, it does
include the energy in the `Coulombic part' of the field associated
with the black hole hair. (Recall that the future limit of the Bondi
energy has this property.) This is all the information that
we need from the IH formalism. In what follows we  
restrict our attention to the family of Static Spherically Symmetric 
(SSS) solutions, as embedded in the IH phase space.

A standard parameterization for the metric and gauge potential
is given by,
\ba
\label{SSS}
\d s^2 &=& -N^2\, e^{-2\delta}\,\d t^2 + N^{-2}\,\d r^2+ r^2\d\Omega^2\, 
,\\ 
\bA &=& a\tau_3\d t +(w\tau_1)\d \theta
+(\cot\theta\tau_3+w\tau_2)\sin\theta\d\phi\, .
\ea
with $N^2=(1-2m/r)$ and
$\delta$,$m$,$a$ and $w$  functions of only  $r$. The (constant) matrices 
$\tau_1,\tau_2$ and $\tau_3$ are the standard basis for $su(2)$.

Roughly speaking, there are two classes of solutions. The first one are
Abelian solutions embedded in SU(2), with electric charge $e$ and
magnetic charge $g$. These are precisely the Reissner-Nordstrom solutions
given by $m(r)=M-(e^2+g^2)/2r$ and $\delta=0$.

The second, and  more interesting sector, corresponds to
non-Abelian solutions to the EYM equation of motion, which are
 known to exist only
for the `magnetic' sector of the theory.
Here we find the regular solitonic solutions\cite{BK} and the
so called colored black hole solutions\cite{gv,bizon,kunzle}.
The curvature takes the standard form \cite{bizon},
\ba
\bF &=& w^\prime\tau_1\,\d r\wedge \d\theta+
w^\prime\tau_2\sin\theta\;\d r\wedge \d\phi\\ \nonumber
&{}&  -(1-w^2)\tau_3\sin\theta\;
\d\theta\wedge\d\phi\, .
\ea
In this case, the equations are known to have a discrete
number of solutions, for each value of the horizon
area, labeled
by an integer $n$ that represents the number of nodes of the function $w(r)$.
The lowest mode, $n=0$, represents the Schwarzschild solution. Therefore,
the solution can be completely parametrized by two numbers 
$(M_{\rm ADM},n)$, the ADM mass and the integer $n$.

{}From an historical perspective, these were the first examples of `hairy
black holes' . They are
`hairy'  because the electric
and magnetic charges defined at infinity
 are both zero, so the only parameter at infinity
is the ADM mass. If the no-hair conjecture were valid for
the EYM system, the
specification of $M_{\rm ADM}$ would suffice to characterize the
solution completely. However, this is not the case, since for a
given value of the ADM mass, there exist a countable number of
{\it different} solutions, labeled by $n$.

Let us now evaluate the HHM for the 
special case of colored black holes.
These solutions are purely magnetic, so $Q_\Delta=0$. The 
formula for the mass (\ref{smarr}) now takes the form,
\be
M_\Delta=\frac{1}{4\pi}\kappa \;a_\Delta + V(r_\Delta,P_\Delta),
\label{12}
\ee
where 
$$\kappa=\frac{e^{-\delta(r_\Delta)}}{2r_\Delta}
\left[1-\frac{P_\Delta^2}{r_\Delta^2}\right]\qquad {\rm and}
\qquad a_\Delta=4\pi r_\Delta^2.$$
(recall that $\delta(\infty)=0$.)
The magnetic charge $P_\Delta$, for the colored black holes 
(\ref{SSS}) is given by:
$P^2_\Delta=(1-w_\Delta^2)^2$.
We expect this formula to reduce to
the Smarr formula for the $n=0$ solution, since in that case
$\delta\equiv 0$ and $w=\pm 1$, so $m_\Delta=r_\Delta/2$. Thus,
for $n=0$, we expect $V\equiv 0$.

The mass variation formula (\ref{11}) coming from the 
isolated horizons framework, when restricted to the purely magnetic
sector of the SSS space takes the form,
\be
\delta M_{\Delta}=\frac{1}{8\pi}\,\kappa\,\delta a_\Delta\, .
\label{22}
\ee
which is the first law  for the HHM.

Now, in order to have a consistent Hamiltonian formulation, 
one should be able to integrate (\ref{22}) to find a function
$M_\Delta$. A detailed
analysis shows that we have a consistent Hamiltonian formulation
if and only if (\ref{12}) and (\ref{22}) are compatible. 
This
in turn implies that $a_\Delta$ and $P_\Delta$ are not free to
vary independently; their variations are constrained to lie in 
one dimensional subspaces of the $(a_\Delta,P_\Delta)$ plane
\cite{cb:ac}. We can then view the magnetic charge 
$P_\Delta$ as a
function of $r_\Delta$.
This fact is, of course, verified in the
explicit numerical solutions reported in the literature.
Next, we can arrive at the condition that the function $V$
should satisfy. If we write 
$\kappa(r_\Delta)=\beta(r_\Delta)/(2r_\Delta)$
it takes the form,
\be
V^\prime=-\frac{r_\Delta}{2}\beta^\prime\, ,
\ee
with `prime' denoting differentiation with respect to $r_\Delta$. 
Furthermore, by requiring that $M_\Delta\mapsto 0$ as
$r_\Delta\mapsto 0$, -coming from
physical considerations- 
we arrive at the following relation,
\be
M_\Delta=\frac{1}{2}\int_0^{r_\Delta} \,\beta(\tilde{r}_\Delta)\,
\d \tilde{r}_\Delta \label{13}
\ee
where the integration is performed over the {\it space of parameters}
of the black hole, labeled by the horizon radius $r_\Delta$, and
not over space-time. This is the first observation of this letter.
Let us note that for the $n=0$ solution, where $\beta$ is known in
closed form ($\beta=1$), we arrive at $M_\Delta^{(n=0)}=r_\Delta/2=
\kappa a_\Delta/(4\pi)$, as expected.
 
Several remarks are in order. First, we must emphasize that the 
determination of $V$, and thus of $M_\Delta$ relied on considerations
involving only variations of quantities associated with the horizon 
$\Delta$. 
Second, the HHM  
defined by (\ref{smarr}), when restricted to SSS configurations,
does not agree with the usual definitions of mass that one finds in
the literature (see for instance \cite{review} and \cite{heusler2}).
It should be stressed that (\ref{13}) comes from a consistent
Hamiltonian formulation, and is {\it not} a definition as occurs
in other treatments.

 There is a general argument from symplectic geometry
that states that, within each connected component of the SSS space
embedded in the space of isolated horizons, the value of the
Hamiltonian $H_t$ remains constant \cite{abf2}. 
Let us review this argument
since it is essential for our discussion.
The Hamilton equations of motion
can be written as $\delta H=\Omega(\delta,X_H)$, where $\Omega$
is the symplectic form, $\delta$
is an arbitrary variation and $X_H$ is the Hamiltonian vector field.
A static solution is one at which the Hamiltonian vector field either
vanishes or generates pure gauge evolution. In either case, 
the symplectic structure evaluated on $X_H$ 
and {\it any} arbitrary vector field 
$\delta$ vanishes. Therefore, for this point of the phase space, 
$\delta H=0$ for any direction $\delta$. In particular $\delta H=0$
for variations relating two static solutions. Now, in the case of
Einstein-Maxwell theory, the no-hair theorems ensure that
all static solutions are given by the RN family. That is, 
the space of static solutions is in that case, connected. Furthermore, 
since there is
no energy scale in the theory, the only possible value for $H_t$
is zero \cite{abf2}. 

What is the situation  in Einstein-Yang-Mills theory? First,
there is the Abelian family of solutions, that represent a connected
component, parametrized by $M,Q,P$. For these solutions, the basic
reasoning of \cite{abf2} applies, with subtle
modifications pertaining to the magnetic solutions \cite{ac:ds}.   
Second, as we mentioned above,
the  EYM system  possesses an energy scale, so in principle,
non-zero values of $H$ are allowed.
Each connected component of the space of SSS
colored black holes is one-dimensional (parametrized by $r_\Delta$),
 and
solutions corresponding to distinct values of $n$ belong to
disconnected  components. 
That is, the space SSS has a countable number of
connected components. As we shall now show,
for $n\geq 1$ the value of the Hamiltonian turns out 
to be {\it different} from zero: $H_t^n\neq 0$. 

Recall that the general argument described above tells us that the 
(on shell) value of the Hamiltonian is constant for each family 
labeled by $n$. This in particular implies that its value is independent of
 the radius $r_\Delta$ of the horizon. Thus one is allowed to take the limit
\be
H^{(n)}=\lim_{r_\Delta\mapsto 0} [M^{(n)}_{\rm ADM}
(r_\Delta)-M^{(n)}_{\Delta}(r_\Delta)]\, .
\ee
Now, it is known that the colored black holes converge point-wise to the
Bartnik-McKinnon soliton solutions \cite{BK} and that the ADM mass satisfies
$M^{(n)}_{\rm ADM}\mapsto M^{(n)}_{\rm BK}$ 
when $r_\Delta \mapsto 0$. Furthermore, 
the mass of the black hole goes to zero
in this limit, so we can  conclude that
\be
H^{(n)}=M^{(n)}_{\rm BK}\, ,
\ee
that is, the total value of the Hamiltonian equals the mass of the $n$th
Bartnik-McKinnon soliton solution!

We now collect our results and arrive at the unexpected relation,
\be
M^{(n)}_{\rm ADM}(r_\Delta)
=M^{(n)}_{\rm BK} + 
M^{(n)}_{\rm \Delta}(r_\Delta)\, .
\ee
Thus, we are in the position of writing
an explicit formula for the ADM mass
of the $n$ colored black hole as function of $r_\Delta$,
\be
M^{(n)}_{\rm ADM}(r_\Delta)
=M^{(n)}_{\rm BK} + 
\frac{1}{2}\int_0^{r_\Delta} \,\beta^{(n)}(\tilde{r}_\Delta)\,
\d \tilde{r}_\Delta\, .
\label{adm2}
\ee
This formula has been numerically tested for the first  
colored black holes, finding complete agreement \cite{ac:ds}.
This is the second observation of this note.
It is important to stress that, {\it a priori},
one would not expect to get the value
of quantities defined at infinity, like the difference of ADM masses
in terms of purely local quantities at $\Delta$.

We can now try to understand the physical meaning of the relation
(\ref{adm2}).
Two facts are known about these solutions:
first, we know that for fixed $a_\Delta$ these solutions represent saddle
points of the ADM mass function $M$ \cite{sud:wal}, and
thus, as one can expect, 
for all values of $n$ these solutions are unstable under small 
perturbations \cite{uns}. 
Let us now note that for the reported solutions in the literature 
(see, for instance, \cite{BK}),
the BK Mass is a monotonic function of $n$, starting at
$M_{\rm BK}^1\approx 0.828$ and approaching $1$ as $n$
grows (in standard normalized units). 
The fact that the mass of the
soliton, and therefore the total energy of the colored
black holes is positive, confirms our expectation,
coming from energetic considerations, that in general
 $M^{\rm ADM} \geq M_\Delta$. 
Furthermore, as the difference between the HHM and ADM  
mass can be seen as the 
energy that is available for radiation to fall both into 
the black hole and to infinity, one can understand the nonzero value of the
Hamiltonian as an indication that there is a potentiality for instability
of the solution. That is, a necessary condition for the solution 
to be unstable is for the value of Hamiltonian on the 
solution in question to be positive. 

Let us end with two remarks. First,
note that the first law for the HHM (\ref{11})
is consistent with
the results given by different formalisms  at infinity, 
where it has been shown that
the ADM mass varies as \cite{sud:wal,heusler2},
\be
\delta M_{\rm ADM}=\frac{1}{8\pi}\,\kappa\,\delta a_\Delta\, ,
\label{20}
\ee
but since we know that
$\delta H=\delta(M_{\rm ADM}-M_{\Delta})=0$,
we have complete agreement with (\ref{22}).
Finally, let us note that
it should be possible to apply the type of analysis presented here to other 
theories where nontrivial static solutions have been found. In particular,
Einstein Yang Mills Higgs, Einstein Yang Mills Dilaton,
and  Einstein Skyrme Theories, are examples in
which there are both solitonic  and Black Hole solutions.

We would like to thank A. Ashtekar, C. Beetle, S. Fairhurst
and U. Nucamendi for discussions,  R. Wald for correspondence, 
and the Center for Gravitational Physics and
Geometry for its hospitality. This work was in part supported by
DGAPA-UNAM grant No IN121298, by a NSF-CONACyT colaborative
grant, by NSF grants INT9722514, PHY95-14240 and by the
Eberly research funds of Penn State.


\begin{thebibliography}{99}



\bibitem{2} J.W. Bardeen, B. Carter  and S.W. Hawking, The four laws
of black hole mechanics \textit{Commun.\ Math. \ Phys.} \textbf{31},
161 (1973).

\bibitem{chru} Chru\'sciel P., No Hair Theorems Folklore, Conjectures,
Results, \textit{Contemporary Mathematics} {\bf 170}, 23 (1994). 

\bibitem{mh} M. Heusler,  \textit{Black Hole Uniqueness Theorems}
(Cambridge University Press, Cambridge, 1996)

\bibitem{bizon2} P. Bizon,  Gravitating solitons and hairy black holes,
\textit{Acta \ Phys.\ Polon.} {\bf B 25}, 877 (1994). 

\bibitem{ack} A. Ashtekar, A. Corichi  and K. Krasnov,  Isolated
Horizons: The Classical Phase Space, \textit{Adv.\ Theor.\ Math.\
Phys.} {\bf 3}, 418 (1999). (Preprint gr-qc/9905089.)


\bibitem{abf2} A. Ashtekar, C. Beetle  and S. Fairhurst, Mechanics of
isolated horizons,  \textit{Class.\ Quantum Grav.} {\bf 17}, 253 (2000). 


\bibitem{ac:ds} A. Corichi, U. Nucamendi and D. Sudarsky, 
Einstein-Yang-Mills
Isolated Horizons: Phase Space, Mechanics, Hair and Conjectures, 
(Preprint gr-qc/0002078).



\bibitem{BK} R. Bartnik and J. McKinnon,  Particlelike Solutions to the
Einstein-Yang-Mills Equations, \textit{Phys.\ Rev.\ Lett.} {\bf 61}, 141
(1988).

\bibitem{gv} M. S. Volkov and D. V. Gal`tsov, Non-Abelian
Einstein-Yang-Mills black holes, \textit{JETP Lett.} {\bf 50}, 346
(1989).

\bibitem{bizon} P. Bizon, Colored Black Holes, \textit{Phys.\
Rev.\ Lett.} {\bf 64}, 2844 (1990).

\bibitem{kunzle} H.P. Kunzle  and A.K.M. Masood-ul-Alam, Spherically
symmetric static SU(2) Einstein-Yang-Mills fields, \textit{J.\ Math.\
Phys.} {\bf 31}, 928 (1990).

\bibitem{cb:ac} C. Beetle and A. Corichi, 
Preprint-CGPG (2000).

\bibitem{review} M. S. Volkov  and D. V. Gal'tsov D V 
Gravitating Non-Abelian
solitons and black holes with Yang-Mills fields, \textit{Physics Reports}
{\bf 319}, 1-83 (1999). 

\bibitem{heusler2} M. Heusler and N. Straumann, The first law of
black hole physics for a class of non-linear matter models, \textit{Class.\
Quantum\ Grav.} {\bf 10}, 1299 (1993).

\bibitem{sud:wal} D. Sudarsky and R. Wald,  Extrema of mass,
stationarity, and staticity, and solutions to the Einstein-Yang-Mills 
equations, \textit{Phys.\ Rev.} {\bf D 46}, 1453 (1992).

\bibitem{uns} N. Straumann and Z. H. Zhou,  Instability of a
colored black hole solution, \textit{Phys.\ Lett.} {\bf B243}, 33
(1990).


\end{thebibliography}
\end{document}